\documentclass[preprint2]{aastex}
\usepackage[english]{babel}
\usepackage{subfigure,epsfig, geometry, times}

\def\msun{M_\odot}
\def\pac{Paczy\'{n}ski}
\def\that{{\hat t}}
\def\umin{u_{\rm min}}
\def\amax{{A_{\rm max}}}
\def\ten#1{\times 10^{#1}}
\def\VEV#1{\left\langle #1\right\rangle}
\newcommand{\be}{\begin{equation}}
\newcommand{\ee}{\end{equation}}
\newcommand{\lab}[1]{\label{#1}}

\newcommand{\chisq}{\chi^2}
\newcommand{\badevent}{}

\begin{document}

\title{Galactic Bulge Microlensing Events from the MACHO Collaboration}

\author{      
      C.L.~Thomas\altaffilmark{1},
      K.~Griest\altaffilmark{1},
      P.~Popowski\altaffilmark{2},
    K.H.~Cook\altaffilmark{3},
    A.J.~Drake\altaffilmark{4},
    D.~Minniti\altaffilmark{4},
    C.~Alcock\altaffilmark{5},
    R.A.~Allsman\altaffilmark{6},
      D.R.~Alves\altaffilmark{7},
    T.S.~Axelrod\altaffilmark{8},
      A.C.~Becker\altaffilmark{9},
    D.P.~Bennett\altaffilmark{10},
    K.C.~Freeman\altaffilmark{11},
      M.~Geha\altaffilmark{12},
    M.J.~Lehner\altaffilmark{13},
    S.L.~Marshall\altaffilmark{14},
    D.G.~Myer\altaffilmark{1},
    C.A.~Nelson\altaffilmark{3},
    B.A.~Peterson\altaffilmark{11},
    P.J.~Quinn\altaffilmark{15},
    C.W.~Stubbs\altaffilmark{5},
      W.~Sutherland\altaffilmark{16},
      T.~Vandehei\altaffilmark{1},
      D.L.~Welch\altaffilmark{17} \\
      (The~MACHO~Collaboration) \\ 
      }

\altaffiltext{1}{Department of Physics, University of California,
    San Diego, CA 92093, USA\\
    Email: {\tt clt, kgriest, dmyer@ucsd.edu, vandehei@astrophys.ucsd.edu }}
 
\altaffiltext{2}{Max-Planck-Institute for Astrophysics,
    Karl-Schwarzschild-Str.\ 1, Postfach 1317, 85741 Garching bei M\"{u}nchen, Germany\\
    Email: {\tt popowski@mpa-garching.mpg.de}}

\altaffiltext{3}{Lawrence Livermore National Laboratory, Livermore, CA
    94550, USA\\
    Email: {\tt kcook, cnelson@igpp.ucllnl.org}}

\altaffiltext{4}{Departmento de Astronomia, Pontifica Universidad Catolica, Casilla 104, Santiago 22, Chile\\
Email: {\tt dante, ajd@astro.puc.cl}}

\altaffiltext{5}{Harvard-Smithsonian Center for Astrophysics, 60 Garden St., Cambridge, MA 02138, USA\\
Email: {\tt calcock, cstubbs@cfa.harvard.edu}}

\altaffiltext{6}{NOAO, 950 North Cherry Ave., Tucson, AZ  85719, USA\\
  Email: {\tt robyn@noao.edu}}

\altaffiltext{7}{Laboratory for Astronommy \& Solar Physics, Goddard Space Flight Center, Code 689, Greenbelt, MD 20781, USA\\
    Email: {\tt alves@astro.columbia.edu}}

\altaffiltext{8}{Steward Observatory, University of Arizona, Tucson,
AZ  85721, USA\\
Email: {\tt taxelrod@as.arizona.edu}}

\altaffiltext{9}{Astronomy Department,
    University of Washington, Seattle, WA 98195, USA\\
    Email: {\tt becker@astro.washington.edu}}

\altaffiltext{10}{Department of Physics, University of Notre Dame, IN 46556, USA\\
    Email: {\tt bennett@emu.phys.nd.edu}}

\altaffiltext{11}{Research School of Astronomy and Astrophysics,
        Canberra, Weston Creek, ACT 2611, Australia\\
 Email: {\tt kcf, peterson@mso.anu.edu.au}}

\altaffiltext{12}{Carnegie Observatories,  813 Santa Barbara Street,
     Pasadena, CA 91101, USA\\
     Email: {\tt mgeha@ociw.edu}}

\altaffiltext{13}{Department of Physics and Astronomy, University of
Pennsylvania, PA 19104, USA\\ 
Email: {\tt mlehner@hep.upenn.edu}}

\altaffiltext{14}{SLAC/KIPAC, 2575 Sand Hill Rd., MS 29, Menlo Park,
CA 94025, USA\\
Email: {\tt marshall@slac.stanford.edu}}

\altaffiltext{15}{European Southern Observatory, Karl-Schwarzchild-Str.\ 2,
        85748 Garching bei M\"{u}nchen, Germany\\	
	Email: {\tt pjq@eso.org}}

\altaffiltext{16}{Institute of Astronomy, University of Cambridge,
Madingley Road, Cambridge. CB3 0HA, U.K.\\
    Email: {\tt wjs@ast.cam.ac.uk}}

\altaffiltext{17}{McMaster University, Hamilton, Ontario Canada L8S 4M1\\
    Email: {\tt welch@physics.mcmaster.ca}}

\begin{abstract}

We present a catalog of 450 high signal-to-noise microlensing events observed by the MACHO collaboration between 1993 and 1999. The events are distributed throughout our fields and, as expected, they show clear concentration toward the Galactic center. 
No optical depth is given for this sample since no blending efficiency 
calculation has been performed, and we find evidence for substantial blending.
In a companion paper we give optical depths for the sub-sample of events on clump giant source stars, where blending is not a significant effect.

Several events with sources that may belong to the Sagittarius dwarf galaxy are identified. For these events even relatively low dispersion spectra could suffice to classify these events as either consistent with Sagittarius membership or as non-Sagittarius sources. 
Several unusual events, such as microlensing of periodic variable 
source stars, binary lens events, and an event showing extended source effects are identified.  We also identify
a number of contaminating background events as cataclysmic variable stars. 


\end{abstract}


\section{Introduction}

The structure and composition of our Galaxy is one of the outstanding problems in contemporary astrophysics.  While inventories of bright stars have been made, it is known that the bulk of the material in our Galaxy is dark.  In addition, the number and mass distribution of stellar remnants such as white dwarfs, neutron stars and black holes is quite uncertain, as is the number of faint stars, brown dwarfs and extra-solar planets.  Gravitational microlensing was suggested as a probe to detect compact objects including dark matter (\pac\ 1986, Griest 1991a, Nemiroff 1991) and was observationally discovered 
in 1993 (Alcock et al.\ 1993; Aubourg et al.\ 1993; Udalski et al.\ 1993).  The line-of-sight towards the Large Magellanic Cloud (LMC) is best for dark matter detection, but as a probe of faint objects from planets to black holes, the line-of-sight towards the Galactic bulge is superior (Griest et al.\ 1991b; \pac\ 1991). The high density of stars in the disk and bulge means that the vast majority of events detected by microlensing experiments are in this direction.

The amount of lensing matter between a source and the observer is typically described using the optical depth to microlensing, $\tau$, defined as the probability that a given source star will be magnified by any lens by more than a factor of 1.34. Early predictions (Griest et al.\ 1991b; \pac\ 1991) of the optical depth towards the Galactic center included only disk lenses and found values near $\tau = 0.5\ten{-6}$. The early detection rate (Udalski et al.\ 1993, 1994a) seemed higher than this, and further calculations (Kiraga \& \pac~1994) added bulge stars to bring the prediction up to $0.85 \ten{-6}$.  The first measurements were substantially higher than this, $\tau \geq 3.3 \pm 1.2 \ten{-6}$ (Udalski et al.\ 1994b) based upon 9 events and $\tau = 3.9^{+1.8}_{-1.2}$ (Alcock et al.\ 1997) based upon an efficiency calculation and 13 clump-giant events taken from their 45 candidates.  
Many additional calculations ensued, including additional effects, especially non-axisymmetric components such as a bar 
(e.g. Zhao, Spergel \& Rich 1995; Metcalf 1995; Zhao \& Mao 1996; Bissantz et al.\ 1997; Gyuk 1999; Nair \& Miralda-Escud\'e 1999; Binney, Bissantz \& Gerhard 2000; Sevenster \& Kalnajs 2001; Evans \& Belokurov 2002; Han \& Gould 2003). 
Values in the range $0.8 \ten{-6}$ to $2 \ten{-6}$ were predicted for various models, and values as large $\tau=4\ten{-6}$ were found to be inconsistent with almost any model.

More recent measurements have all used efficiency calculations and have found values of $\tau = 3.2 \pm 0.5 \ten{-6}$ from 99 events in 8 fields using difference imaging (Alcock et al.\ 2000b), $\tau = 2.0 \pm 0.4 \ten{-6}$ from around 50 clump-giant events in a preliminary version of the companion paper (Popowski et al.\ 2001a), $\tau = 0.94 \pm 0.29 \ten{-6}$ from 16 clump-giant events (Afonso et al.\ 2003), and 
$\tau = 3.36^{+1.11}_{-0.81}$ 
from 28 events (Sumi et al.\ 2003). 

In releases similar in spirit to our catalog here Udalski et al.\ (2000) presented a catalog of 214 microlensing events from the 3 seasons of the OGLE-II bulge observation, and Wo\'zniak et al.\ (2001) presented a catalog of 520 events, mainly from difference imaging. 

In this work we present our complete catalog of high signal-to-noise microlensing events that were found with point spread function fitting photometry. In the companion paper (Popowski et al.\ 2004), we make an accurate determination of the bulge optical depth using the 62 clump giant events (60 unique) listed here and find $\tau = 2.17^{+0.47}_{-0.38} \ten{-6}$ at $(l,b) = (1 \hbox{$.\!\!^\circ$} 50, -2 \hbox{$.\!\!^\circ$} 68)$. We do not calculate an optical depth for our entire sample of microlensing events since a complete blending efficiency calculation has not been performed, and we caution against using the entire sample of events for such purposes.

Initially envisioned as a probe of dark matter, microlensing has evolved into a more general astronomical tool, useful for several distinct purposes. For example, since the duration of the microlensing event is proportional to the square root of the lens mass, microlensing is sensitive to compact objects in the $10^{-7}\msun$ to $10^1\msun$ range, independent of the object's luminosity, so it facilitates inventories of brown dwarfs, white dwarfs, and black holes.  However, the lens mass measurement is degenerate with the lens distance and speed, severely limiting the accuracy of the mass distribution measurement.  Our catalog includes several long duration events that may be massive black holes and several short duration events that may be brown dwarfs.

In order to break the mass/distance/speed degeneracy several techniques have been applied to rare classes of events such as those with binary lenses, binary sources, large annual parallaxes, etc.  Our catalog lists events which may be members of exotic microlensing classes.
Finally microlensing has emerged as a powerful method of detecting or constraining the existence of extra-solar planets orbiting the lens (Mao \& \pac~1991; 
Gould \& Loeb 1992; Griest \& Safizadeh 1998; Rhie et al.\ 2000; Gaudi et al.\ 2002). 
A key for these searches is careful follow-up on microlensing events, almost all of which are towards the Galactic bulge.  Our catalog can be used to determine the frequencies of detectable lensing in various directions towards the bulge.

For comparison with other works we note that our definition of microlensing
event duration, $\that$, is the Einstein ring diameter crossing time, 
twice the more commonly used Einstein ring radius crossing time.  

\section{Data}

\begin{figure}[t]
\epsfig{file=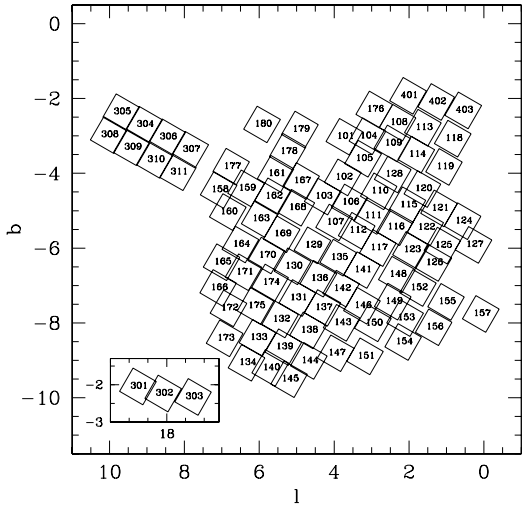,width=0.48\textwidth}
\caption{\label{fig:field_locs} Location of the bulge fields in galactic coordinates}
\end{figure}

The MACHO Project had full-time use of the 1.27 meter telescope at Mount Stromlo Observatory, Australia\footnote{A fire tragically razed Mount Stromlo Observatory in January of 2003.} from July 1992 until December 1999.  Details of the telescope system are given by Hart et al.\ (1996), and details of the camera system by Stubbs et al.\ (1993) and Marshall et al.\ (1994).  Briefly, corrective optics and a dichroic were used to give simultaneous imaging of a 43'$\times$ 43' field in two non-standard filter bands, using eight $2048^2$ pixel CCD's.  
A total of 32700 exposures were taken in 94 fields towards the Milky Way bulge, 
resulting in around 3 
Tbytes of raw image data and photometry on 50.2 million stars. 
The location of the centers of the bulge fields are shown in Figure $ \ref{fig:field_locs} $ and the location and number of exposures taken of each field are given in Table $\ref{tab:fielddat}$. Table $\ref{tab:fielddat}$ 
also gives the number of stars in each field, the number of clump giants, the number of microlensing events, and the sampling efficiency at event durations
of $\that=50$ and $\that=200$ days.  
This latter numbers can be used as a rough indication of the relative sensitivity to microlensing in each field, but 
should be used for quantitative work only with the clump giant sample 
of events (see the companion paper by Popowski et al.\ 2004).  
The coverage of fields varies greatly from 12 observations of field 106 to 1815 observations of field 119.  Note that the observing strategy changed several times during the project, so even in a given field the frequency of observations changed from year to year.  In addition, there are gaps between November and February as the bulge was not observed during prime LMC observing times.

The photometric reduction used here is a variation of the DOPHOT (Schechter, Mateo, \& Saha 1993) 
point spread function (PSF) fitting method (Alcock et al.\ 1999). Briefly, a good-quality image of each field is chosen as a template and used to generate a list of stellar positions and magnitudes.  The templates are used to ``warm-start'' all subsequent photometric reductions, and for each star we record information on the flux, an error estimate, the object type, the $\chi^2$ of the PSF fit, a crowding parameter, a local sky level, and the fraction of the star's flux rejected due to bad pixels and cosmic rays.  The resulting data are reorganized into lightcurves, and searched for variable stars and microlensing events. The photometric data base used here is about 450 
Gbytes in size. We report magnitudes using a global, chunk-uncorrected photometric relations that express Johnsons $V$ and Kron-Cousins $R$ in terms of the MACHO intrinsic magnitudes $b_M$ and $r_M$ as:
\begin{equation}
V = b_M - 0.18(b_M-r_M) + 23.70
\end{equation}
\begin{equation}
R = r_M + 0.18(b_M-r_M) + 23.41.
\end{equation}
For more details see Alcock et al.\ (1999).

\begin{figure}[t]
\epsfig{file=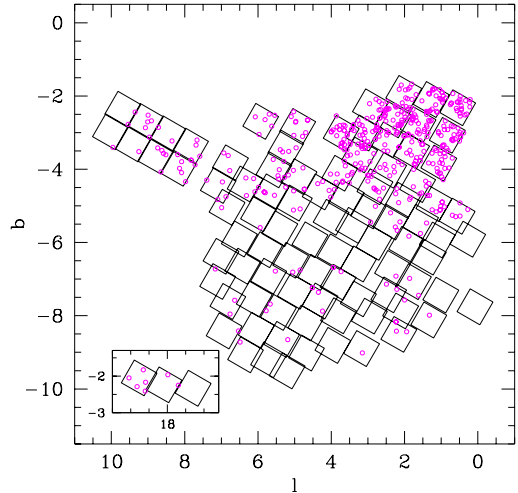,width=0.47\textwidth}
\caption{\label{fig:event_dist} Spatial distribution of microlensing events.}
\end{figure}

\section{Event selection}

The data set used here consists of about 19 billion individual photometric measurements.  Discriminating genuine microlensing from stellar variability, systematic photometry errors, and other astronomical events is difficult, and the significance of the results depend upon the event selection criteria.  The selection criteria are based on cuts made on a set of over 150 statistics calculated for each lightcurve. 
First a smaller set of ``level 1'' statistics is calculated for every star in the data base.  Based on variability criteria, a few percent of the lightcurves are advanced to level 1 and the complete set of statistics, including non-linear fits to microlensing lightcurve shape, are calculated. Using these a broad selection of events is advanced to level 1.5 and output.
Finally a fine-tuned set of selection criteria is used to select the level 2 
candidates. From a total of 50.2 million lightcurves, around 90000 were advanced to level 1.5, and 337 to level 2. In addition, during this procedure, lightcurves are tagged as variable stars for inclusion in our variable star catalog.

If the goal is to measure the optical depth or microlensing rate, then great care must taken in the selection of candidate microlensing events. It is crucial that the same selection method be performed on the actual data and on the artificial data used to calculate the detection efficiency. A certain fraction of ``good" microlensing events will be missed by any set of selection criteria, and it is important to not include these events in any calculation of optical depth. For more discussion, the reader is referred to the companion paper (Popowski et al.\ 2004) that derives the microlensing optical depth toward the Galactic bulge based on clump giant events.  

The basic cuts of selection criteria C are given in Table \ref{tab:cuts}. A more thorough description of the most useful statistics is given in Alcock et al.\ (2000c).  Note that Table \ref{tab:cuts} below is not intended to fully inform about the selection of bulge events; instead it is supposed to document the level 2 selection criteria used. Nevertheless in column 2 we offer a rough guide to what a given cut or class of cuts is intended to achieve.

In this paper, however, since no estimate of optical depth is being made, we can augment the computer selected (criteria C) events with ``good" events found by other methods such as our alert system (Alcock et al.\ 1996) or even a search by eye over a larger set of candidates.  
In Figure~\ref{fig:event_dist} we show the position of the microlensing
events on the sky.
In Figure~\ref{fig:lightcurves} we display the lightcurves of events
we selected, organized by field.tile.seq identification number.  
In Tables \ref{tab:table_of_events} and \ref{tab:table_of_nonevents}
we describe the source stars corresponding to the selected events, and also give the microlensing fit parameters. 
Table \ref{tab:table_of_events} contains the events we subjectively regard
as likely microlensing candidates, while Table \ref{tab:table_of_nonevents}
shows the events we think are probably not microlensing.
Column 10 of Tables~\ref{tab:table_of_events} and \ref{tab:table_of_nonevents} shows our subjective ``grade" of the event data quality (A-F). Column 11 shows the method or methods by which each event was selected (`c' $\mapsto$ ``criteria C", `a' $\mapsto$ ``alert system", `b' $\mapsto$ ``binary search", `e' $\mapsto$ ``by eye selection" 
out of an expanded set of events from a `C'-like selection).  
Column 12 shows our subjective determination of event 
type (`CV' $\mapsto$ suspected cataclysmic variable or supernova, 
`var' $\mapsto$ suspected variable star, 
`bin' $\mapsto$ suspected binary lensing event, 
`$R \neq B$' $\mapsto$ red and blue lightcurves differ in shape and/or amplitude indicating a possible 
blend or systematic error). 
In column 1 we also mark events identified as lensed clump giants in the companion paper with a $\dagger$ flag.  We also include OGLE event identifications in Table~\ref{tab:matches} for events found at the same position in both surveys.

Note, that for the clump giant subsample used to calculate optical depth it is important that very few non-microlensing events are selected by the ``C" criteria. However, when we apply the ``C" criteria to non-clump areas of the color-magnitude diagram a few non-microlensing events are selected. This is not a problem for optical depth calculation but may be of interest. We list the entire set of events that pass criteria C, including 5 events which we subjectively graded as probably non-microlensing (quality D or F or suspected variable) in Table \ref{tab:table_of_nonevents}.
  

In summary we show the lightcurves of 252 grade ``A" (very good 
signal-to-noise) candidate microlensing events, 198 grade ``B" (good quality)
 events, and 76 grade ``C'' (poor quality) events.  Of the grade ``A'' events 220 were selected with criteria C, 6 from the alert system, 4 from our binary search, and 22 by eye.  Of the grade ``B'' events 97 were found with criteria C, 15 with the alert system, 2 from the binary search, and 84 by eye.  Of the grade ``C'' events 11 were selected with criteria C, 12 by the alert system, 1 from our binary search, and 52 by eye.  We also identify 32 pairs
 of candidates at the same location on the sky and one triplet of events. 
These are either the same physical events reported in two overlapping fields (14 cases) or two stars so close on the sky that they both receive the
 flux from the actual event (20 cases). This latter effect results from the photometry code and not the microlensing of two separate sources.  In such cases we move the worse of the two events into Table \ref{tab:table_of_nonevents}, and recommend ignoring it.

\section{Special events}
\subsection{Binaries}

Alcock et al.\ (2000a) described 17 binaries in the Galactic bulge. We
include these events in the lightcurve figures and in the tables. We also mark 24 additional events as potential binaries.  These are
events that have deviations from the standard lightcurve shape and may be
better fit with a binary lens or source lightcurve.  We have not done this
fitting in this paper, and it is also possible that these are not
microlensing or have larger than normal measurement errors.

\subsection{Lensing of variable stars}

Several of the good quality microlensing events occurred on periodic, or nearly
periodic variable stars.  These include events:
108.18689.1979, 108.19602.415, 118.18009.35, and 403.47848.35.

These events are useful because the measured amplitude of the stellar variation allows one to determine the amount of blending.  If the variability can be used to learn more about these stars (such as their distance or radius) then in some cases the degeneracy between lens mass, distance, and velocity may be partly broken. 
\subsection{Other exotic events}
Event 121.22423.1032 seems to display extended source effects.

\section{Supernovae and Cataclysmic Variables}

Supernova (SN) explosions in galaxies behind the microlensing source stars
have been shown to contribute a significant background to the LMC and
SMC microlensing searches (Alcock et al.\ 2000c).  
We do not expect SN to be as important
in this search towards the Galactic bulge due to the large extinction
through the disk and bulge, but we did a search for SN-like lightcurves,
and have marked a number of events that we think are not microlensing.
In fact, most of these events are probably cataclysmic variables (CV), e.g., novae or dwarf novae (DNe), so we mark them as `CV'.  
Of the 16 events we identify in this way, 7 repeat, i.e. show more
than one brightening.  Most of these events exhibit a rapid rise ($\sim 4$ days)
followed by a more gradual decline ($\sim 20$ days).  The peak is typically
about 4 magnitudes brighter in $V$ and 2.7 magnitudes brighter in $R$ than the baseline, consistent
with the CV classification (Sterken \& Jaschek, 1996).
We identify these events as CV's rather than SNe because the decline after
peak is too fast over the first 20 days as compared with typical SN.

The lightcurves of the repeating CV's
can be seen in Figure \ref{fig:lightcurves}.
We classify these as long period dwarf novae, since the periods seem to be 
between 300 and 700 days.  
In particular note:
event 113.18676.5195 with 7 outbursts and a period of around 400 days,
event 114.19842.2283 with 5 outbursts and a period of around 340 days,
event 115.22695.3361 with 3 outbursts, as well events with two outbursts:
178.23266.2918, 178.24048.3166, and 311.37730.4143. 

Since our photometry points are generally separated by at least
one day, no flickering analysis is done.  In dwarf novae
one expects flickering on time scales of minutes to hours, so further
observations are needed to positively identify these source as DNe.
The 9 events with single excursions are more difficult to identify;
possibilities include long duration DNe, or heavily blended 
classical novae.  They are unlikely to be SNe.

\section{The significance of blending}

\lab{sec:blending}
The photometry code measures the light coming from stars within the seeing disk, and for bulge stars and conditions at Mt.\ Stromlo Observatory this means there are usually many stars contained within each photometric ``object".  However, in almost all cases only the light from one of these stars is lensed and gives rise to the transient microlensing lightcurve. The light from the non-lensed source therefore ``blends" with the light from the lensed source distorting the lightcurve from its theoretical shape. In particular the event duration $\that$ derived from a fit to a blended lightcurve can be shortened and the magnification
decreased.  
Since the microlensing optical depth is proportional to the durations of the events, blending must be taken into account when trying to measure an optical depth.

In the companion paper (Popowski et al.\ 2004), 
we show that when using clump giant stars as sources the problem of blending is much alleviated.  In the events on non-clump giant stars listed in this paper, however, blending is expected to be quite significant. One signature of a heavily blended event is a large difference between the magnification in the red and blue filter bands.  In Tables \ref{tab:table_of_events} and \ref{tab:table_of_nonevents}
we label events which have such a large difference as ``$R\neq B$".  
These differences may be due to blending, or especially for the lower quality events (grade C) these differences may just be indicating that the event is not microlensing.  Because of this effect it is important to use only the clump giants for any quantitative work.

\section{Signatures of Sagittarius dwarf galaxy}

Sagittarius dwarf galaxy (Ibata, Gilmore, \& Irwin 1995) is the
closest satellite galaxy to the Milky Way at a distance of about 25 kpc
from the Sun. It is centered at the globular cluster M54 
at $(l,b) = (5 \hbox{$.\!\!^\circ$} 6, -14 \hbox{$.\!\!^\circ$} 0)$,
and extends over several
degrees perpendicular to the Galactic plane.
Traces of Sgr dwarf structure can be seen behind the MACHO
fields that lie at negative Galactic latitudes. Therefore, we
expect that some microlensing events may originate in Sgr dwarf.
The identification of events with sources in the Sgr dwarf
serves several goals: 1. it removes the contaminating sources from
the map of the microlensing optical depth toward the Galactic bulge
and thus improves the determination of bar parameters; 2.  it
probes the inner 25 kpc of the Galaxy for massive dark structures; 3. it helps to constrain the mass function of the lenses.  We discuss these points in more detail below.

To fully explore the results of the microlensing surveys,
we would like to better understand the lens population. In particular,
we want to assign the lenses to different Galactic
populations. However, because most of the lenses are too faint to be directly
observed, we attempt to use the location of the sources to constrain
the location of the lenses. We can assume that the sources in the
Galactic bulge imply that the lenses are either in the bulge or in
the disk and that the sources in the Sagittarius dwarf galaxy should
typically have lenses in the bulge.  By finding events that have Sgr
sources, we can make better maps of the microlensing optical depth
toward the sources in the bulge.  Such improved maps will provide
crucial constraints in constructing better models of the Galactic
bar.  It is even possible that a detailed analysis of events with Sgr
sources could reveal a lens population {\em behind} the Galactic
bulge. Such a population could be part of the warped or flared disk
or even of a new, as yet undiscovered streamer of stars. In brief,
Sgr events probe the inner 25 pc of the Galaxy for intervening
structures in a way not possible with microlensing events with sources
in the bulge.

A separate goal is to constrain the masses of the lenses.  The
distribution of the durations of events contains the information
about the masses of the lenses and the kinematics of the objects
involved in the lensing process. However, a characteristic time of
microlensing event is a degenerate combination of several parameters,
including the geometry of the system and the relative transverse
velocity.  The better constraints we have on the kinematics, the
better we can understand the masses of the lenses. For example, Gould
(2000) showed that the bulge velocity dispersion introduces so much
scatter to the duration distribution that lenses in the form of brown
dwarfs cannot be distinguished from those in the form of neutron
stars on an event-by-event basis.  Therefore, the situations where kinematics is additionally
constrained are very valuable.  There are a few generic cases that
help to determine the masses of the lenses: 1) the parallax effect,
which places constraints on the combination of the relative velocity
and distances (Bennett et al.\ 2002), 2) the measurement of the
relative proper motion of the lens with respect to the source, which
is particularly powerful if coupled with a parallax measurement
(Alcock et al.\ 2001), 3) the possibility to assign a source to a
system with distinct bulk velocity and negligible velocity dispersion
(e.g., Sagittarius dwarf galaxy). We think the time is ripe to explore this
third option. The extent to which the identification of the
microlensing events with sources in Sagittarius dwarf galaxy would
improve the determination of the masses of the lenses can be judged
from Fig. 8 by Cseresnjes \& Alard (2001). Moreover, such events can probe a
different lens population than all the other techniques used thus
far. Cases 1) and 2) are biased
toward detecting the lenses in the disk, whereas the lenses for Sgr
events would likely be in the Galactic bulge or may even be behind
the bulge.

\begin{figure*}[p]
\begin{center}
\epsfig{file=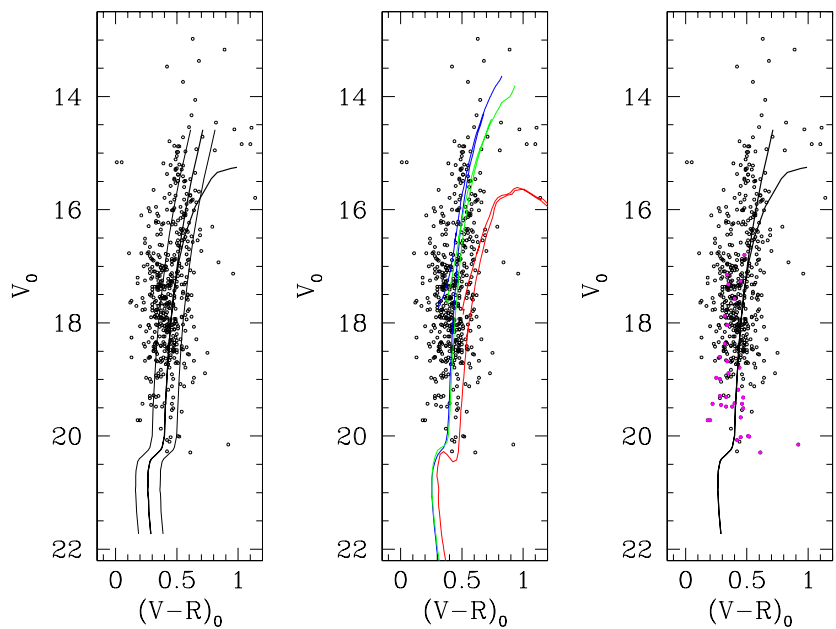, angle=-90, width=.8\textwidth}
\end{center}
\caption{
In panel a) we plot the events together with a
ridge-line (bold lines) of the Sgr dwarf galaxy taken from
Bellazzini et al.\ (1999). 
The intrinsic width of the observed
stellar distribution in Sgr dwarf is visualized with thin solid lines
(only for the bluer branch of the Bellazzini et al.\ 1999 track).  
The errors of $(V-R)_0$ color of the MACHO events are of order of at
least 0.05 mag.
In panel b) we superpose relevant Padova isochrones on the same collection
of events. The blue isochrone is for an age of 12.6 Gyr and metallicity
[M/H] of $-1.7$, the green one for an age of 10.0 Gyr and metallicity $-1.3$,
and the red one for an age of 6.3 Gyr and metallicity of $-0.4$.
We shifted the isochrones assuming the distance modulus, 
$(m-M)_{\rm Sgr} = 17.0$.
From panels a) and b) we conclude that many microlensing events are 
consistent with having
sources in the Sgr dwarf galaxy.
Therefore, the location of events on the $(V,V-R)$ color-magnitude
diagram does not facilitate the identification of Sgr sources.
In panel c), the events with the Galactic latitude $b<-6.0$ are
marked as solid magenta dots. Their distribution on a CMD is not identical to
the other events and many of them are more consistent with Sgr
membership. Again we over-plotted a ridge-line from Bellazzini et al.\ (1999) for reference.
\label{fig:sag1}}
\end{figure*}

Is it possible to select Sgr events based on 
their expected location on a
color-magnitude diagram (CMD) from the MACHO survey?  This is illustrated in 
Figure~\ref{fig:sag1}, where
we show the sources of microlensing events detected by the MACHO
collaboration on a $V_0$ versus $(V-R)_0$ CMD.
The CMD was dereddened using the extinction map by Popowski, Cook, \&
Becker (2003) [extinction $A_V$ was taken from column 4 of their Table 3]. 
In panel a) we plot the microlensing events together with a
ridge-line (bold lines) of the Sgr dwarf galaxy taken from
Bellazzini et al.\ (1999). 
The ridge-line has been adjusted to the dereddened quantities using
$E(V-I) = 0.22$ and $A_V = 0.55$. The ridge-line in $(V-R)_0$ color
has been derived assuming that $(V-R)_0 = 0.5 \, (V-I)_0$, which
according to Padova isochrones (Girardi et al.\ 2002:
the tables provided on their web page: {\tt
http://pleiadi.pd.astro.it})
is accurate to within a few percent.
The intrinsic width of the observed
stellar distribution in Sgr dwarf is visualized with thin solid lines
(only for the bluer branch of the Bellazzini et al.\ 1999 track).  
The errors of $(V-R)_0$ color of the MACHO events are of order of 
at least 0.05 mag. We
conclude that many microlensing events could have
sources in Sgr dwarf galaxy.  
In panel b) we superpose relevant Padova isochrones on the same collection
of events. The blue isochrone is for an age of 12.6 Gyr and metallicity
[M/H] of $-1.7$, the green one for an age of 10.0 Gyr and metallicity $-1.3$,
and the red one for an age of 6.3 Gyr and metallicity of $-0.4$
(which is claimed to be the dominant population according to 
Monaco et al.\ 2002).
We shifted the isochrones assuming the distance modulus, 
$(m-M)_{\rm Sgr} = 17.0$.
Again, many microlensing events are consistent with having
sources in the Sgr dwarf galaxy.
Therefore, the location of events on the $(V,V-R)$ color-magnitude
diagram does not facilitate the identification of Sgr sources.

\begin{figure}[t]
\epsfig{file=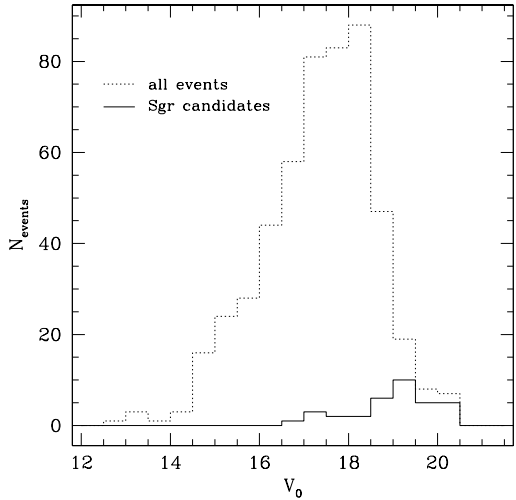, width=.48\textwidth}
\caption{The histogram of $V_0$ magnitudes for a set of all
511 unique microlensing events and 34 Sagittarius candidates.
The difference in magnitude distribution is very significant.
\label{fig:sgr2}}
\end{figure}

Is there any way to narrow the list of possible Sgr events?  Kunder,
Popowski, \& Cook (2004, in preparation) analyzed a set of almost
4000 RR Lyrae stars in the MACHO bulge fields. They separated the
stars into the bulge and Sgr groups with high confidence.  The Sgr
RR Lyrae stars dominate over the bulge ones for Galactic latitudes
$b<-6.0$. This suggests that Sgr sources can make a detectable
contribution to the microlensing optical depth at these latitudes,
which is in qualitative agreement with conclusions from Cseresnjes \& Alard (2001). 
In panel c) of Figure~\ref{fig:sag1}, 
the microlensing events with the Galactic latitude $b<-6.0$ are
marked as solid magenta dots. Their distribution on a CMD is not identical to
the other events, which is apparent from the distribution of their
dereddened $V_0$ magnitudes (Figure~\ref{fig:sgr2}). In addition, many
of those events are in the vicinity of the Sgr ridgeline suggesting
that they are more consistent with Sgr membership. These 34 events are 
our Sgr dwarf candidates.
We list their main parameters in Table~\ref{tab:sag1}.

The Sagittarius dwarf galaxy has a distinct heliocentric
radial velocity of $140 \pm 10$ km/s, different from the bulk of
bulge stars (see e.g. Figure 4 by Ibata et al.\ 1995).
The Sgr membership cannot be assigned in a robust way
based on the measurement of radial
velocities alone, but such measurements are very powerful in
eliminating bulge or disk events. In addition, radial velocity
can be obtained long after the
event. 
Our 34 candidates are the recommended targets for such an
investigation\footnote{Ideally, one
would like to perform such a radial velocity test for all
microlensing candidates from all microlensing surveys, especially the ones 
with negative
Galactic latitude $b$.  Due to the location of the MACHO fields, the
MACHO data are the most suitable for the search for events with Sgr
sources.}.

\begin{figure}[t]
\epsfig{file=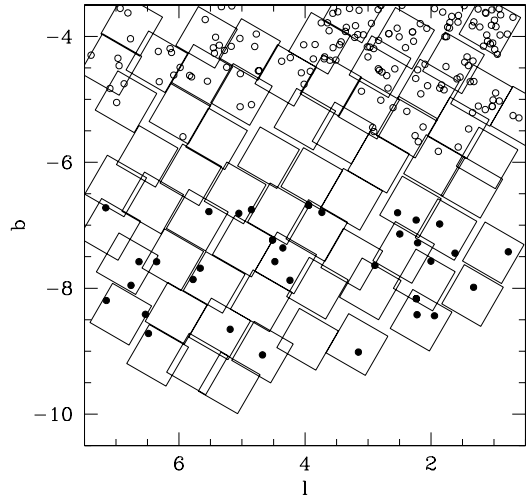, width=.48\textwidth}
\caption{Spatial distribution of events (dots) in the
MACHO fields (squares) most distant from the Galactic plane. The
filled dots represent Sgr candidates. The strip of empty fields at
$b \approx -6$ is caused by very low detection efficiency in those
fields and unrelated to the existence of different event populations.
\label{fig:sag3}}
\end{figure}

Our selection of Sgr microlensing candidates should enable the first
systematic search for such stars by the means of radial velocities. 
There are two recent spectroscopic
studies of microlensing events toward the Galactic bulge by Cavallo
et al.\ (2002) and Kane \& Sahu (2003). The first investigates 6
and the second 17 events. Most of the events studied by those
groups are rather bright and neither of the above studies
specifically targeted the microlensing sources in the Sgr dwarf
galaxy.
An example of 136.27650.2370/142.27650.6057, 
which is not a Sgr member, shows that 
spectroscopic follow-up is essential. 
Event 136.27650.2370/142.27650.6057 is in our candidate list but was 
measured by
Cavallo et al.\ (2002) to have the radial velocity of $60 \pm 2 {\rm
km \, s}^{-1}$, inconsistent with the velocity of Sgr dwarf.
On the other hand, there may be Sagittarius events hiding among
bulge events closer to the Galactic plane. Cook et al.\ (2004) claim
a detection of two likely Sgr events that are distinct through their
radial velocity, metallicity and location on the $(K, J-K)$ CMD.
Determination of radial velocities of our candidate Sgr events
asks for observations on an 8m class telescope.
Unfortunately, these observations cannot be sped up with currently
available multi-object
instruments, because the candidate Sgr events are
distributed over a large area.
Their spatial distribution is shown in Figure~\ref{fig:sag3}.

As many as five methods to identify Sagittarius events are discussed 
by Popowski (2004)\footnote{See: {\tt http://www.stelab.nagoya-u.ac.jp/hawaii/Popowski/hawaii2004.html}.}.

\section{Statistical properties of the events}

\subsection{Clustering of microlensing events}

Figure \ref{fig:event_dist} shows the position of the events on the sky.  
The events are noticably concentrated toward the Galactic center and toward 
the Galactic plane as expected.  
Examination of the figure shows some apparent clustering of events on the sky,
in particular in fields 108, 104, and 113.
If microlensing events are clustered on the sky above random chance
it has important consequences.  It could indicate clustering of lenses,
perhaps in some bound Galactic substructure.
We tested for the significance of the clustering in our data by
simulating 10000 microlensing experiments each of which found 318 criteria ``C'' selected unique microlensing events (as in the current data set).
The Monte Carlo is described in more detail in the companion paper
(Popowski et al.\ 2004).  The result is that we find no strong evidence 
for clustering beyond random chance.  The probability of finding by chance
a cluster of 3 events as dense as in the data is betwen 7 and 36\% depending on the assumed optical depth gradient.  The chance of finding
a 4-event cluster as dense as in the data varies between 4 and 32\% depending on the assumed optical depth gradient.

\subsection{Impact Parameters}
One test of microlensing is the distribution of impact parameters, $\umin$.  
The impact parameter $u_{\rm min}$ is the distance of closest approach between 
the lens and the source in units of the Einstein ring radius, and it 
is completely determined by the maximum magnification. 
If the efficiency were independent of the magnification one would expect 
a uniform distribution of $\umin$'s since every impact parameter 
is equally likely. 
In that case, a cumulative distribution of
impact parameters should be a straight line from 0 up to the maximum impact
parameter allowed by our cuts (the cut $A_{max} >= 1.5$ corresponds to
$\umin<0.826$.)  
\begin{figure*}[t]
\subfigure[]{\epsfig{file=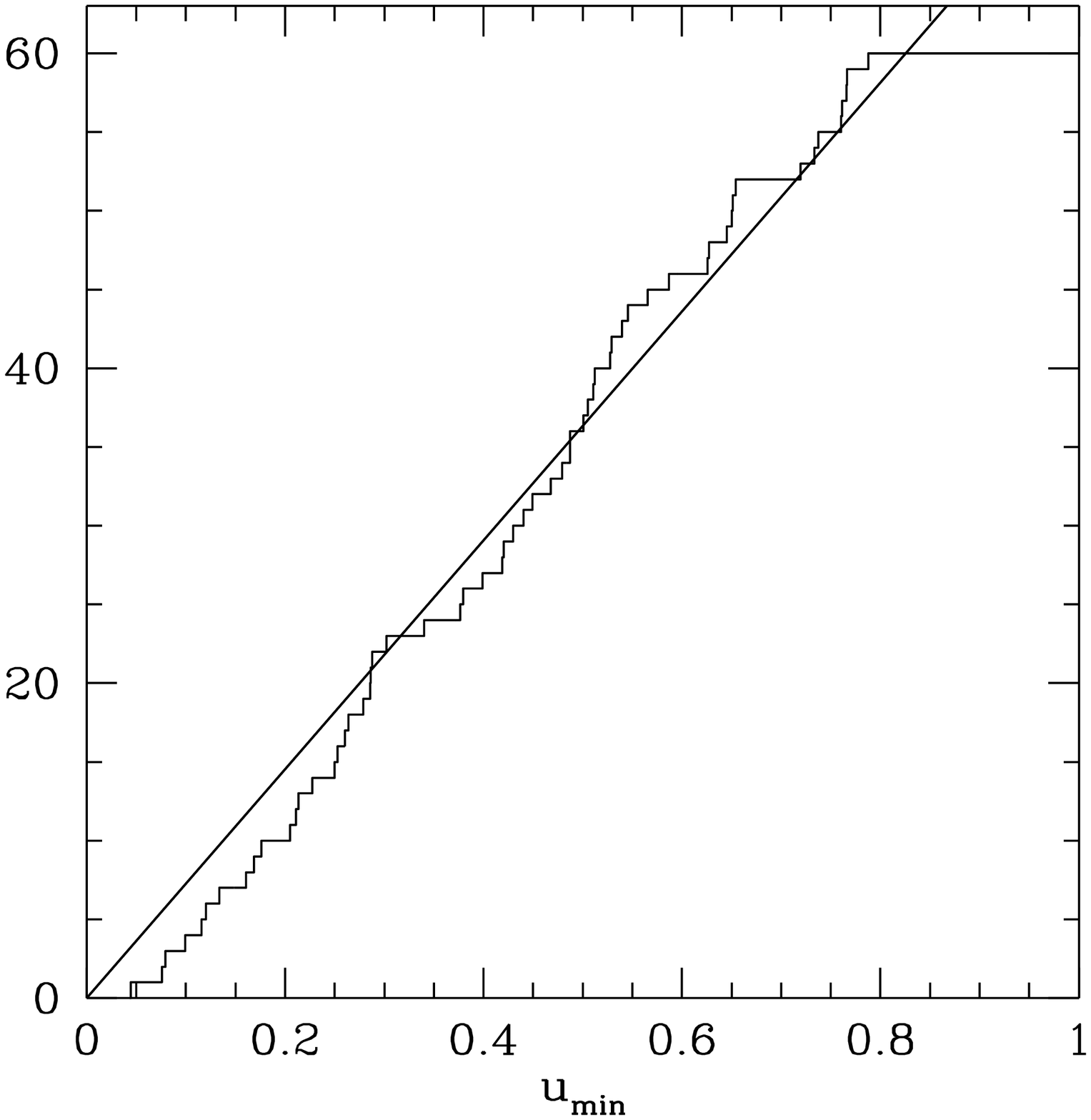,width=.48\textwidth}}
\subfigure[]{\epsfig{file=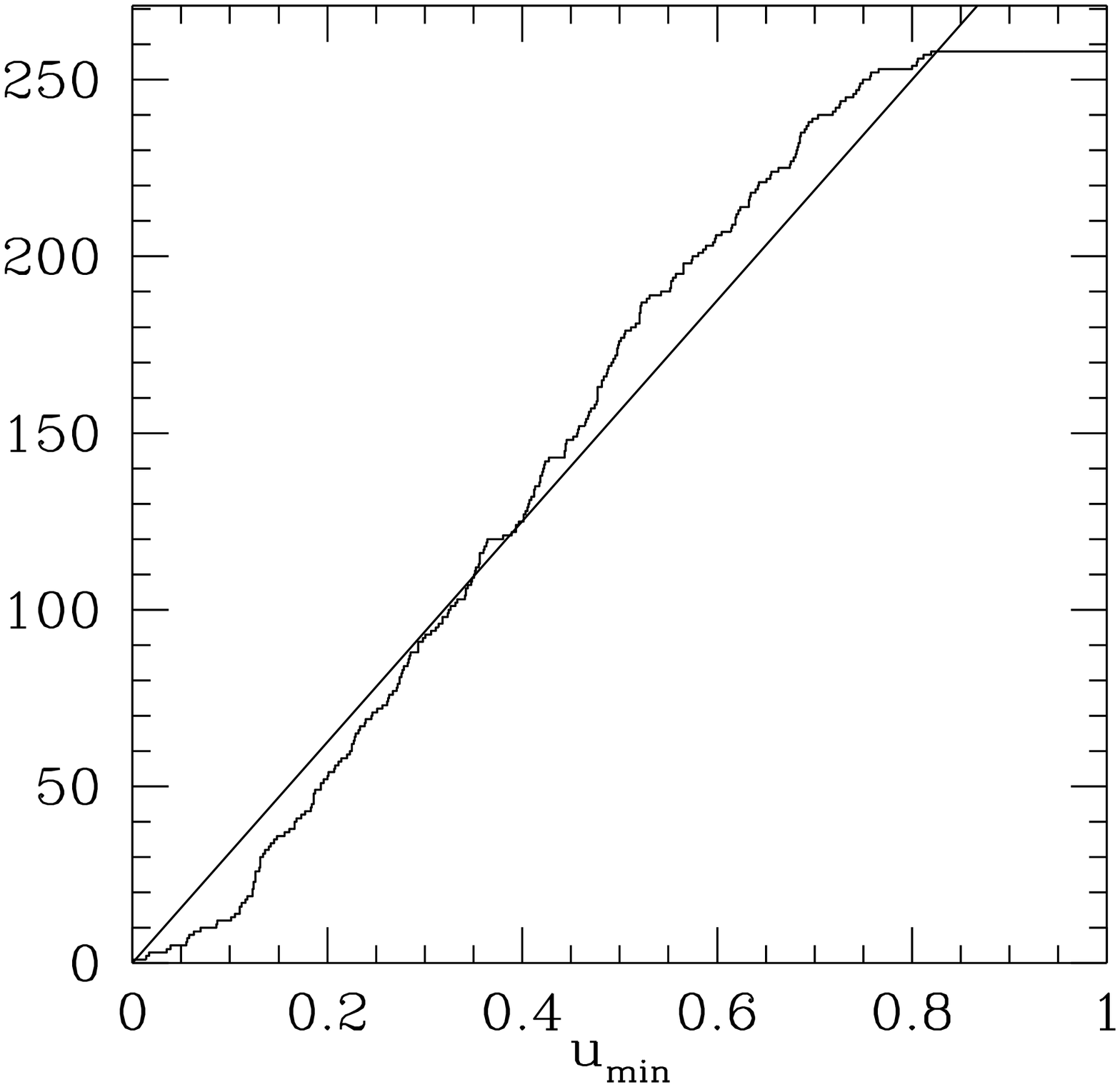,width=.48\textwidth}}
\caption{Cumulative distribution of impact parameter for clump events (a) and non-clump, selection criteria C events (b).  \label{fig:impact_params}}
\end{figure*}
In Figure \ref{fig:impact_params}, we plot the cumulative distributions
of impact parameters for unique events selected by the `C' criteria
for both clump giants events (60 events) and non-clump events (258 events).  
In both cases no correction is made for microlensing
efficiency, though this correction is made (with little effect) for the clump giant events in the companion paper (Popowski et al.\ 2004).

For the clump events, 
the resulting Kolmogorov-Smirnov (KS) statistic shows excellent agreement 
with the microlensing hypothesis:
$D=0.081$ with 60 events, with a probability of 81\% to find a value of $D$ this large or
larger.
For the non-clump events, the agreement is marginal:
$D=0.091$ with 258 events, for a probability of 2.5\% of finding a value of $D$ this large.
This deviation from uniformity can be caused by blending (which can lower
the measured maximum amplification and therefore increase the measured $\umin$), by a lower efficiency at larger impact parameter, or by inclusion 
of non-microlensing events in the sample.  

\subsection{Distributions}

\begin{figure*}[th]
\centering
\mbox{\subfigure[]{\epsfig{figure=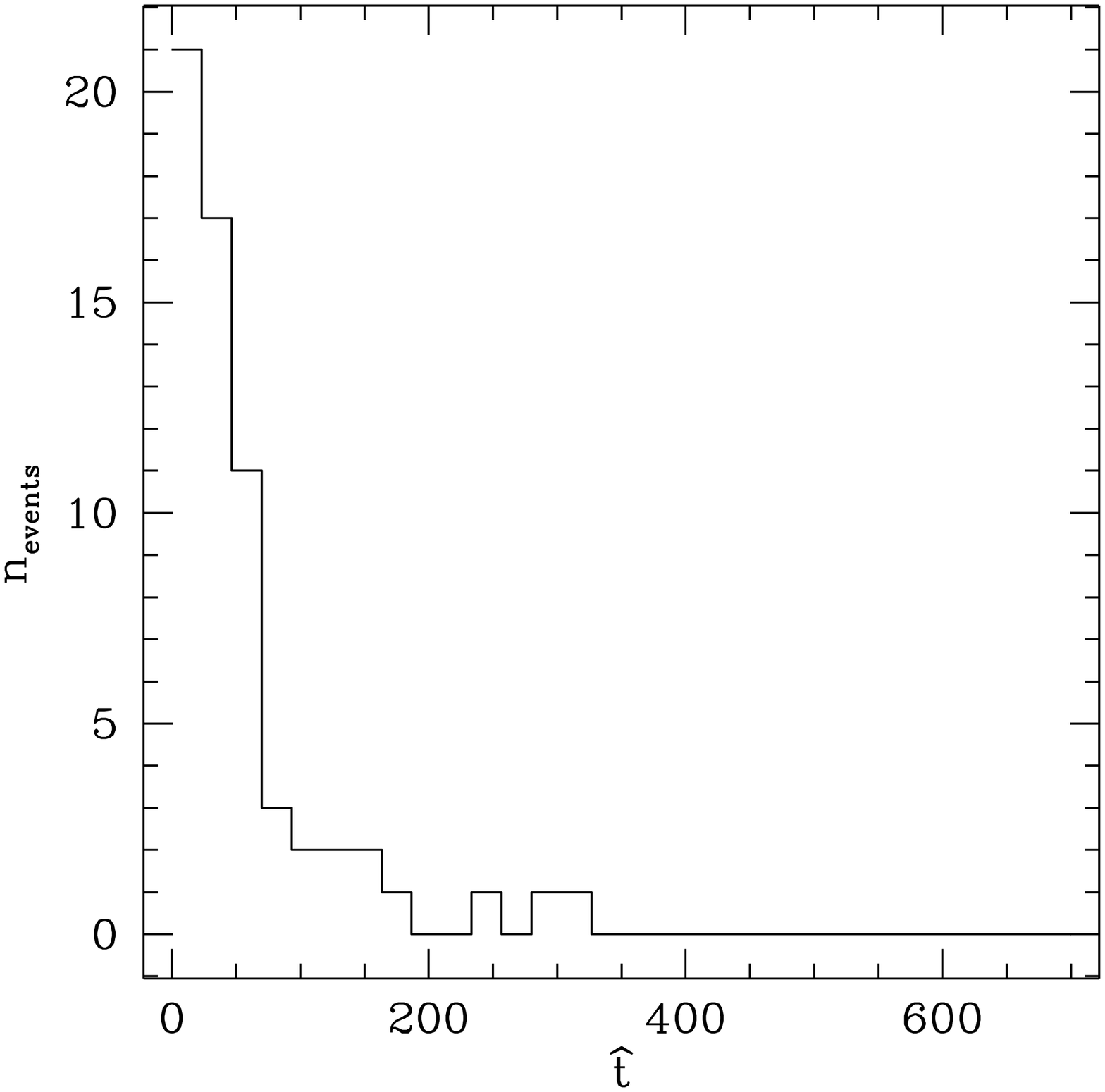,width=0.47\textwidth}}\quad
      \subfigure[]{\epsfig{figure=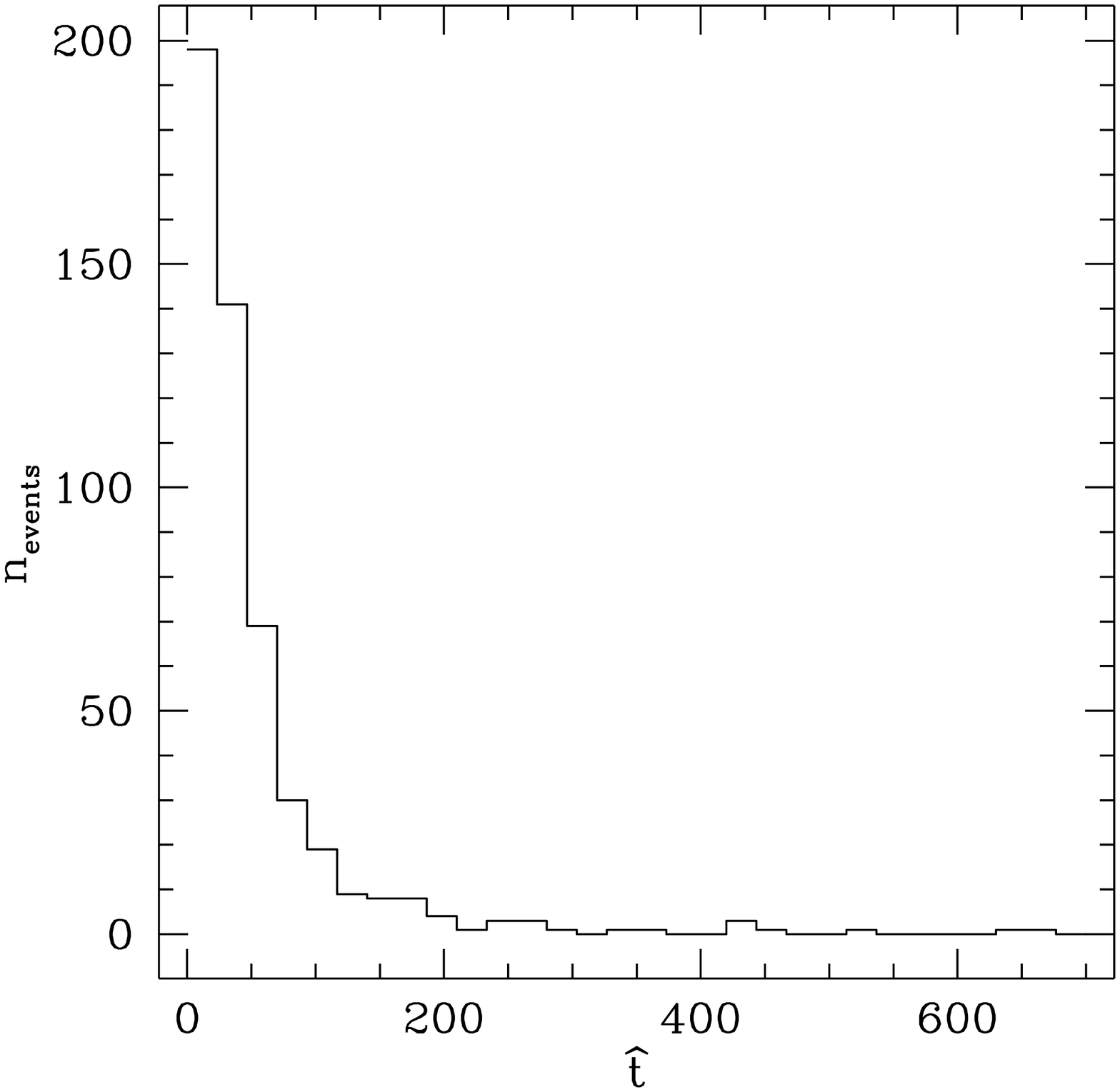, width=0.47\textwidth}}}
\caption{Distribution of event durations for clumps giants (a) and non-clump giants (b). \label{fig:that_dists}}
\end{figure*}

Figure \ref{fig:that_dists} shows the distribution of lensing durations $\that$ for both the clump giant and non-clump giant events.  
Only events grade A and B events are included.
The average value of $\that$ for the non-clump sample is 
$\VEV{\that} = 49 \pm 62$ days.
For comparison note that the clump giant events have 
$\VEV{\that} = 56 \pm 64$ days.  Because the distributions are not gaussian we also give the median and quartiles for non-clump A and B events 31.1, 17.4, \& 57.0, and clump events 30.8, 15.9, \& 60.9.  These results are consistent with partial blending of the non-clump sample discussed in \S~\ref{sec:blending} but they do not provide any additional support for this hypothesis.  

\begin{figure}
\epsfig{figure=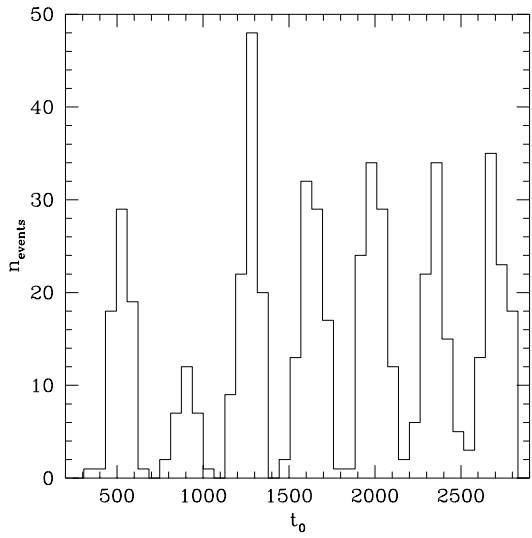, width=0.47\textwidth}
\caption{\label{fig:t0_dist} Histogram of $ t_0 $ of events.  The peaks and troughs are due to the lack of observations form November through February.  The small number of events in the second year is due to fewer observations in that period.}
\end{figure}

\begin{figure}[th]
\epsfig{figure=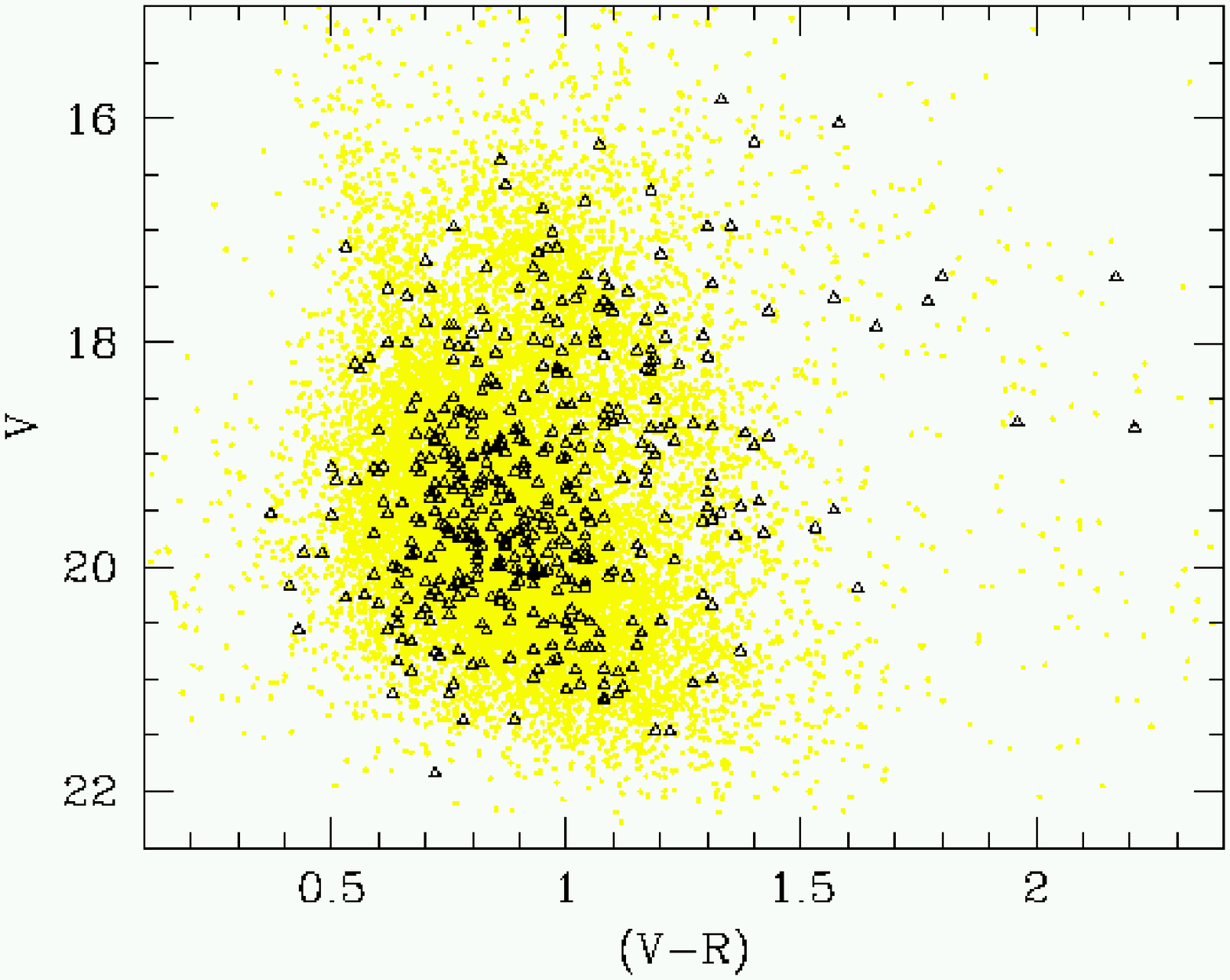, width=0.48\textwidth}
\caption{\label{fig:cmd} Color-magnitude diagram of events (black triangles) and a representative sample of all stars in our fields (yellow dots).}
\end{figure}

Figure \ref{fig:t0_dist} shows the distribution of the times of microlensing peaks, mostly showing when observations took place, but is consistent with uniformity when this is taken into account.  Figure \ref{fig:cmd} shows the CMD of the microlensing events, which, as expected, is a reasonable sample of the CMD of the Galactic bulge.  The location on a CMD is used in the companion paper to select clump giants. 

\section{Conclusions}
In conclusion, light curves and parameters of 528 microlensing events found by the Macho Project (1993-1999) are presented.  Included are 5 events on variable
stars, 17 binary events, 24 potential binary events, and 1 extended source event.  Also included is a representative sample of 36 contaminant events, consisting of 16 cataclysmic variables, and 20 duplicate events.  In addition we select 37 (34 unique) events that are potentially lensed Sagittarius sources.  The sample of over 500 events presented here is effected significantly by blending and should not be used for quantitative studies.  We present light curves for all 564 microlensing and non-microlensing events.  Data and figures will be  available at {\tt http://wwwmacho.mcmaster.ca} upon acceptance of this paper.

\acknowledgments

This work was performed under the auspices of the U.S. Department of
Energy, National Nuclear Security Administration by the University of
California, Lawrence Livermore National Laboratory under contract No.
W-7405-Eng-48. KG and CT were supported in part by the DoE under grant
DEFG0390ER40546.  DM is supported by FONDAP Center for Astrophysics 15010003.

\begin{deluxetable}{ccccccccc}
\tabletypesize{\footnotesize}
\tablecaption{Data on the 94 bulge fields.\label{tab:fielddat}}
\tablehead{\colhead{ field } & \colhead{$ l $}  & \colhead{$b$}  & \colhead{$\frac{n_{stars}}{100}$}  & \colhead{$\frac{n_{clumps}}{100}$}  & \colhead{$n_{events}$}  & \colhead{$n_{exposures}$}  & \colhead{$\epsilon(50days.)$}  & \colhead{$\epsilon(200days)$}}
\startdata
101 & 3.73 & $-$3.02 & 6284 & 629.5 & 30 & 804 & 0.41 & 0.61 \\
102 & 3.77 & $-$4.11 & 6545 & 444 & 15 & 421 & 0.24 & 0.46 \\
103 & 4.31 & $-$4.62 & 6349 & 368 & 5 & 331 & 0.21 & 0.36 \\
104 & 3.11 & $-$3.01 & 5813 & 652 & 28 & 1639 & 0.40 & 0.59 \\
105 & 3.23 & $-$3.61 & 6375 & 539 & 21 & 640 & 0.39 & 0.52 \\
106 & 3.59 & $-$4.78 & 6648 & 346 & 0 & 12 & $ <0.003 $ & $ <0.003 $ \\
107 & 4.00 & $-$5.31 & 6159 & 269 & 0 & 51 & 0.007 & $ <0.003 $ \\
108 & 2.30 & $-$2.65 & 6498 & 790 & 32 & 1031 & 0.49 & 0.75 \\
\enddata
\tablecomments{Efficiencies (columns 8 and 9) are averaged over $\amax$.  The classification of stars as clump giants is expected to be very difficult in 300 series fields, so we refrain from quoting specific numbers for these fields.}
\end{deluxetable}

\begin{deluxetable}{ll}
\tablewidth{0in}
\tablecaption{Selection Criteria\label{tab:cuts}}
\tablehead{ \colhead{Selection} & \colhead{Description} }
\startdata
\multicolumn{2}{l}{{\bf microlensing fit parameter cuts:}} \\
$ N_{(V-R)}>0 $ & require color information \\
$ \chi^2_{out} < 3.0, \; \chi^2_{out} > 0.0 $ & require high quality baselines \\
$N_{amp} \geq 8 $ & require 8 points in the amplified region \\
$N_{rising} \geq 1, \; N_{falling}\geq 1$ & require at least one point in the rising and falling part of the peak \\
$A_{max} > 1.5 $ & magnification threshold \\
($A_{max}-1) > 2.0(\sigma_R + \sigma_B)$ & signal to noise cut on amplification \\
$ \!\!\! \begin{array}{l}
(\sigma_R+\sigma_B) < (0.05\,\delta\chisq)/(N_{amp}\,\chisq_{in}\,\chisq_{out})  \\
(N_{amp}\,\chisq_{out}\,f_{chrom})/(\mathit{befaft}\,\delta\chisq)<0.0003  \\
(N_{amp}\,\chisq_{in})/(\mathit{befaft}\,\delta\chisq)<0.00004 
\end{array} $
&
$ \left\} \raisebox{1.8ex}[4.6ex]{} \right.$ remove spurious photometric signals caused by nearby saturated stars \\
$\delta\chisq/fc2 > 320.0 $ & good overall fit to microlensing lc shape \\
$\delta\chisq/\chisq_{peak} \geq 400.0 $ & same quality fit in peak as whole lc\\ 
$0.5(rcrda+bcrda) \leq 143.0 $ & source star not too crowded \\
$\xi^{auto}_B/\xi^{auto}_R <2.0 $ & remove long period variables \\
$t_0 > 419.0,\, t_0 < 2850.0 $ & constrain the peak to period of observations \\
$\hat{t} < 1700 $ & limit event duration to $\sim$half the span of observations \\
\hline
\multicolumn{2}{l}{{\bf clump giant cuts:}}\\
$V \geq 15,\, V\leq 20.5 $ &  select bright stars with reliabale photometry \\
$V \geq 4.2(V-R)+12.4 $ & define bright boundary of extinction strip\\
$V \leq 4.2(V-R)+14.2 $ & define faint boundry of extinction strip  \\
$(V-R) \geq(V-R)_{boundary} $ & avoid main sequence contamination \\
exclude fields $300-311$ & avoid disk contamination \\
\enddata
\end{deluxetable}%


\clearpage

\begin{figure*}[p]
\epsfig{figure=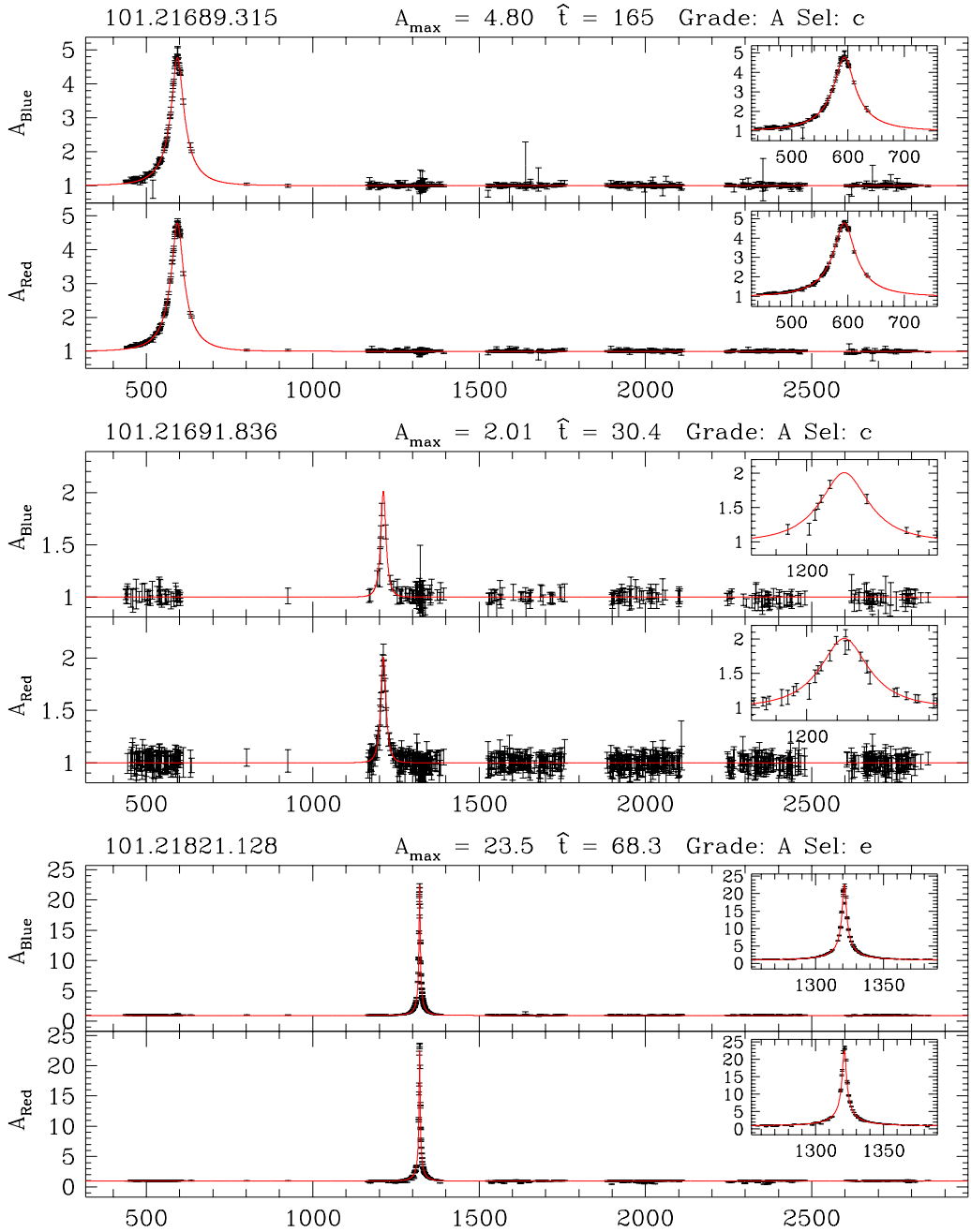, width=0.95\textwidth}
\caption{\label{fig:lightcurves} Example lightcurves of microlensing candidates and some non-microlensing candidates, organized by field.tile.sequence.  Also given are $A_{max}$, $\hat{t}$, our subjective grade of data quality (A-F), the method by which the event was selected, and any notes concerning our classification of the event.  All lightcurve data and figures will be available at {\tt http://wwwmacho.mcmaster.ca}.}
\end{figure*}

\end{document}